\documentclass{cpbtex}

\usepackage{graphicx}
\usepackage{mathrsfs}
\usepackage{caption}
\usepackage{latexsym}
\usepackage{CJK}
\usepackage{dcolumn}
\usepackage{bm}
\usepackage{float}
\usepackage{color}

\definecolor{red}{RGB}{209,75,78}
\definecolor{yellow}{RGB}{230,155,3}
\definecolor{green}{RGB}{101,147,74}

\begin{document}

\title{Simulated Bifurcation Algorithm for MIMO Detection}

\author{\ Wen Zhang$^{1}$\thanks{Corresponding author. E-mail:zhangwen20@huawei.com}, and Yu-Lin Zheng$^{1}$\\
$^{1}${Hisilicon research, Huawei Technologies Co., Ltd., Shenzhen, China}}  

\date{}
\maketitle

\begin{abstract}
We study the performance of the simulated bifurcation (SB) algorithm for signal detection in multiple-input multiple-output (MIMO) system, a problem of key interest in modern wireless communication systems. Our results show that SB algorithm can achieve significant performance improvement over the widely used linear minimum-mean square error decoder in terms of the bit error rate versus the signal-to-noise ratio, as well as performance improvement over the coherent Ising machine based MIMO detection method.
\end{abstract}



\section{Introduction}
Optimal maximum likelihood MIMO (ML-MIMO) detection is an NP-hard problem. So ML-MIMO detector like sphere decoder is impractical to be implemented for large number of antennas because of exponential computational complexity. Suboptimal linear decoders with polynomial complexity, such as minimum-mean square error (MMSE) decoder, enable near-optimal bit error rate (BER) performance in Massive MIMO regime, in which the number of antennas is much larger than the number of users. While the BER performance is poor in large MIMO system where the number of antennas is equal to that of users.

The ML-MIMO detection problem can be formulated into a quadratic unconstrained binary optimization (QUBO) problem or Ising model problem~\cite{ref1}. Some heuristic optimization methods based on Ising model solver, such as quantum annealer~\cite{ref1}, simulated annealing~\cite{ref2}, oscillator Ising machine~\cite{ref3} and coherent Ising machine (CIM)~\cite{ref4}, has been proposed to solve the ML-MIMO detection problem recently and achieve near-optimal solutions with low computational cost. We propose a near-optimal ML-MIMO detector based on simulated bifurcation algorithm and achieve better BER performance than above methods and also linear decoder MMSE.
Ising model and simulated bifurcation

\section{Ising model and simulated bifurcation}

The Hamiltonian or the cost function of Ising model is defined by $H_{Ising}=\sum^N_{i,j}J_{ij}s_is_j+\sum^N_ih_is_i$, where spins $s_i\in{-1,1}$.Simulated bifurcation algorithm simulates adiabatic evolution of a classical nonlinear Hamiltonian system which contains Hamiltonian of an Ising model and approximates the optimal solution of corresponding Ising model~\cite{5}. The nonlinear system Hamiltonian and the evolution described by equation of motions of variables are given by
\begin{equation}
	H_{SB}=\frac{a_0}{2}\sum^N_iy_i^2+\frac{a_0-a(t)}{2}\sum^N_ix_i^2-\frac{c_0}{2}\sum^N_{i,j}J_{ij}x_ix_j+\sum^N_ih_ix_i,
\end{equation}
\begin{equation}
	\dot{x}_i=\frac{\partial H_{SB}}{\partial y_i} = a_0y_i,
\end{equation}
\begin{equation}
	\dot{y}_i=-\frac{\partial H_{SB}}{\partial x_i} = -(a_0-a(t))y_i-c_0(\sum_jJ_{ij}\mathrm{sgn} (x_j)+\frac{1}{2}h_i),
\end{equation}
where the spins $s_i$ in Ising model Hamiltonian are substituted into continuous variables $x_i$. In the evolution $x_i$ is replaced by $\mathrm{sgn}(x_i)$ and set $y_i=0$  when $|x_i|>1$. In the end of the evolution the sign of $x_i$ gives an solution of the Ising model as $s_i=\mathrm{sgn}(x_i)$. $a_0$ is a constant and is set to $1$ for simplicity, $c_0=1/(2\sqrt{N}\lambda)$ with $\lambda = \sqrt{\frac{\sum_{ij}J_{ij}^2}{N(N-1)}}$ guarantees the variables bifurcate to an sub-optimal solution of Ising model, and $a(t)$ is a time dependent function to conduct the adiabatic evolution schedule and is set as a linear function from $0$ to $1$. The sign function in the second term is introduced by digital simulated bifurcation~\cite{ref6} to reduce the computational cost and suppress the discretization error. 

\section{ML-MIMO detection and its Ising formulation}

Consider an uplink $N_t\times N_r$ MIMO system with $N_t$ users and $N_r$ antennas and channel transmission matrix $\mathbf{H}\in \mathit{C}^(N_t\times N_r)$. The ML-MIMO detection problem is to search the transmitted symbols that minimizes the error with respect to ideally-received signal symbol

\begin{equation}
	\mathbf{x}_{ML}=\mathop{\mathrm{argmin}}\limits_{\mathbf{x}\in\Omega^N_t}\|\mathbf{y}-\mathbf{Hx}\|^2,
\end{equation}
where $\mathbf{x}$ and $\mathbf{y}=\mathbf{Hx}+\mathbf{n}$ are the transmit vector and received vector with white Gaussian noise(AWGN) $\mathbf{n}$ respectively, and the elements $x_i$ and $y_i$ are symbol transmitted by user $i$ and symbol received by antenna $j$ respectively. And each $x_i$ is a complex number drawn from a constellation $\Omega$ corresponding to modulation.

The Ising formulation of ML-MIMO detection problem can be found in~\cite{ref1}. Here we briefly introduce the formulation for BPSK, QPSK, and 16-QAM constellation. For QPSK and 16-QAM constellation, complex valuedH, y and x should be transformed to be real valued as

\begin{equation}
	\tilde{\mathbf{x}}=
	\begin{bmatrix}
	\mathit{Re}(\mathbf{x})\\
	\mathit{Im}(\mathbf{x})
	\end{bmatrix},
	\tilde{\mathbf{y}}=
	\begin{bmatrix}
	\mathit{Re}(\mathbf{y})\\
	\mathit{Im}(\mathbf{y})
	\end{bmatrix},
	\tilde{\mathbf{H}}=
	\begin{bmatrix}
	\mathit{Re}(\mathbf{H}) & -\mathit{Im}(\mathbf{H})\\
	\mathit{Im}(\mathbf{H}) & \mathit{Re}(\mathbf{H})
	\end{bmatrix},
\end{equation}

Each element of $\tilde{\mathbf{x}}$ takes integral values in the range ${-bps+1,-bps+3,\dots,bps+1}$and should be expressed by $\mathrm{log}_2bps$ bits, where $bps$ means bits per symbol. Then we can use  $2N_t  \mathrm{log}_2bps\times 1=N_tbps\times 1$  spins vector s for QPSK and 16-QAM to express $\tilde{\mathbf{x}}$. Ising model Hamiltonian can be described by vector form as $\mathbf{H}_{Ising}=\mathbf{s}^\mathrm{T} \mathbf{Js}+\mathbf{hs}^\mathrm{T}$. In this form, the coefficients in Ising model of ML-MIMO detection problem should be

\begin{equation}
	\mathbf{J}=
	\left\{
	\begin{aligned}
	zeroDiag(\tilde{\mathbf{H}}^\mathrm{T}\tilde{\mathbf{H}}) & , & \mathrm{for\quad QPSK}, \\
	zeroDiag(\tilde{\mathbf{H}}\mathbf{T}^\mathrm{T}\tilde{\mathbf{H}}\mathbf{T}) & , & \mathrm{for\quad 16QAM}
	\end{aligned}
	\right.
\end{equation}

\begin{equation}
	\mathrm{with}\quad \mathbf{T} = 
	 \begin{bmatrix}
	 \mathbf{I}(N_t) & \mathbf{I}(N_t) & 0 & 0\\
	 0 & 0 & 2\mathbf{I}(N_t) & 2\mathbf{I}(N_t) 
	 \end{bmatrix},
\end{equation}

\begin{equation}
	\mathbf{h}=
	\left\{
	\begin{aligned}
	-2\tilde{\mathbf{y}}^\mathrm{T}\tilde{\mathbf{H}} & , & \mathrm{for\quad QPSK}, \\
	-2\tilde{\mathbf{y}}^\mathrm{T}\tilde{\mathbf{H}}\mathbf{T} & , & \mathrm{for\quad 16QAM}
	\end{aligned}
	\right.
\end{equation}

\section{Results and discussion}
\subsection{Performance of simulated bifurcation on large MIMO detection}
\begin{figure}[htbp]
\centering
    \includegraphics[width = 0.8\textwidth]{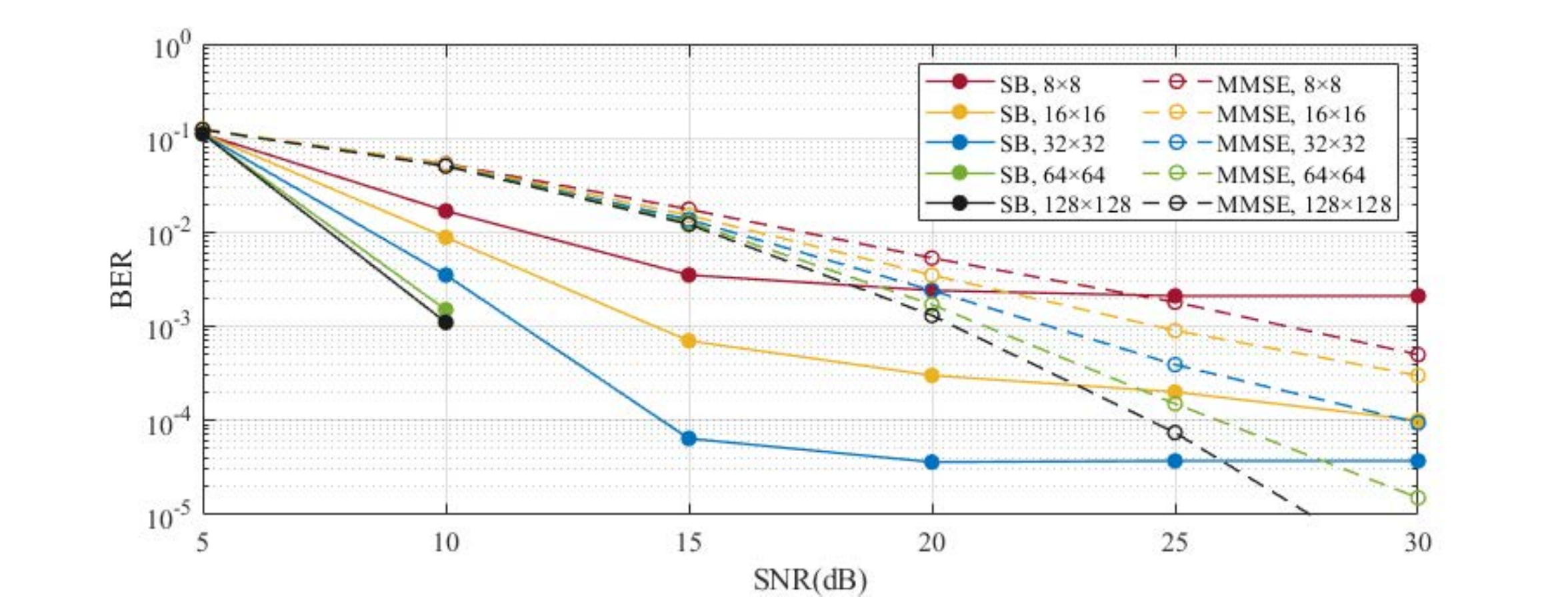}
    \caption{BER performance of SB vs. MMSE on MIMO system with different sizes and QPSK.}
    \label{fig1}
\end{figure}

We evaluate the BER performance of simulated bifurcation on MIMO systems for QPSK modulation with different $N_t\times N_r$ and different different signal-to-noise ratio (SNR). For each scenario, 10,000 random instances are simulated to evaluate the BER, that is, the BER is averaged on $N_t\times20,000$ bits. The number of steps of time evolution in the simulated bifurcation algorithm is set to 100. This provides a good balance between the latency and BER performance of MIMO detection. Figure~\ref{fig1} shows the BER curves of SB and MMSE decoder for different sizes of MIMO systems with QPSK modulation. The missing data point of SB decoder in large size MIMO systems is due to that no errors are found in a limited number of instances we simulated. The results show that simulated bifurcation has a significant performance advantage over the MMSE decoder, especially in the range of 10-15 dB SNR, and this performance advantage becomes more significant as the MIMO size increases. In the case of very high SNR, the BER of SB decoder does not decrease continuously as that of the MMSE decoder. This is caused by the error floor phenomenon encountered by the Ising model solver~\cite{ref4}.

\subsection{Performance of simulated bifurcation with regularization}
\begin{figure}[htbp]
\centering
    \includegraphics[width = 0.8\textwidth]{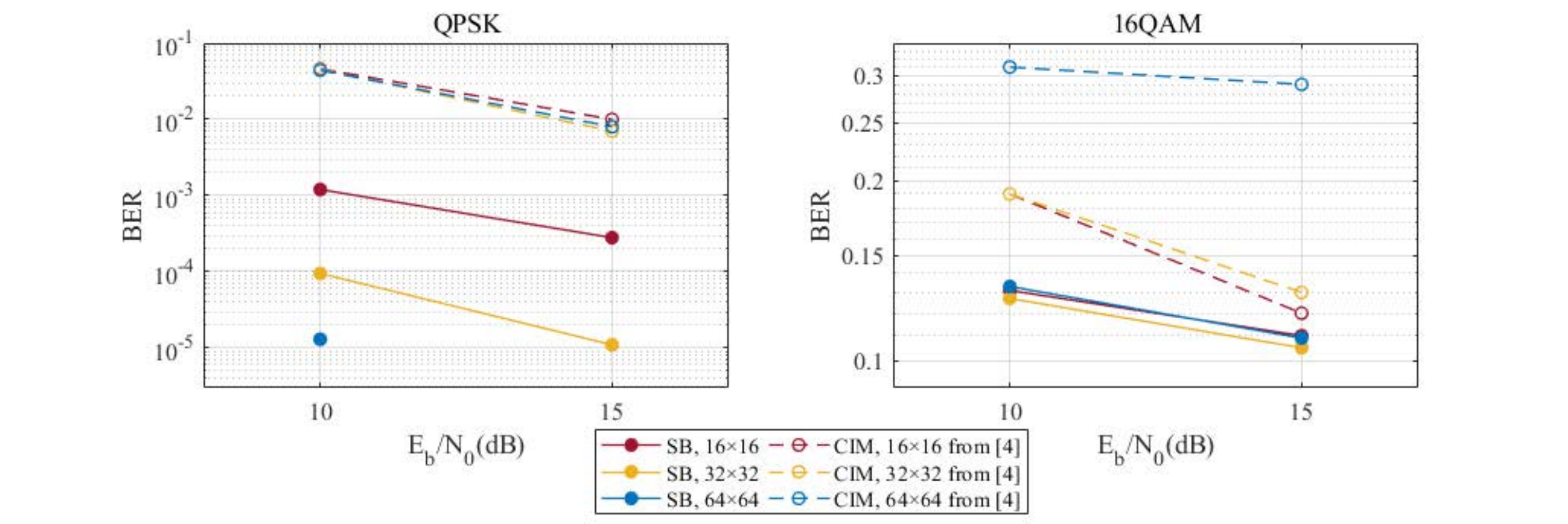}
    \caption{BER performance of SB vs. CIM with regularization on MIMO system with different sizes and modulations.}
    \label{fig2}
\end{figure}

To mitigate the error floor of Ising model solver,\cite{ref4} proposes a regularization approach, which introduces a penalty term to the Ising model Hamiltonian of ML-MIMO:
\begin{equation}
	\mathbf{H}_{Ising-regu} = \mathbf{s}^\mathrm{T}\mathbf{Js}+\mathbf{hs}\mathrm{T}+r\|\mathbf{s}-\mathbf{s}_p\|^2
\end{equation}
where $\mathbf{s}_p$ is Ising spin vector transformed from a suboptimal solution obtained by some linear detector such as MMSE, and $r$ is the regularization parameter. It can force the Ising model to approach the optimal solution from the suboptimal solution in the case of high SNR. Finally, the solution with lower Hamiltonian of Ising model is selected among the solutions of Ising model solver and MMSE. This approach of introducing regularization also applies to simulated bifurcation. We evaluate the BER performance of simulated bifurcation with MMSE solution as a regularization term and compare it with that of CIM with regularization and the results are shown in Figure~\ref{fig2}. The BER data of CIM decoder with regularization is captured from~\cite{ref4} by choosing the lowest BER with optimal r, while the BER of SB with regularization is evaluated with fixed $r=0.5$. The number of instances used to evaluate BER is 10,000 for QPSK and 1,000 for 16QAM respectively. It is shown that SB with regularization has a significant BER performance advantage over CIM on MIMO with difference sizes for QPSK. For 16QAM, the above advantage still exists even though BER becomes worse than QPSK due to higher order modulation.

\subsection{Conclusion}
We explore that the performance of simulated bifurcation algorithm on ML-MIMO detection with various sizes and modulation. Our results show that SB can achieve significant performance improvement over linear decoder MMSE and other Ising solver CIM. This indicates show that the simulated bifurcation is a very competitive detection approach for MIMO systems.



\end{document}